\documentclass{appolb}
\usepackage{epsfig,psfig}
\usepackage{cite}

\def\eq#1{(\ref{#1})}
\newcommand{\Imag}{\mathrm{Im}\,}
\newcommand{\be}{\begin{equation}}
\newcommand{\ee}{\end{equation}}
\newcommand{\ba}{\begin{eqnarray}}
\newcommand{\ea}{\end{eqnarray}}
\begin{document}
\title{Triangle Anomaly and the Muon $g-2$\,\footnote{Based on talks
    given by AC or AV 
    at the XLIII Cracow School of Theoretical
    Physics (Zakopane, May 2003), Conference on the Intersections of
    Particle and Nuclear Physics (New York, May 2003), and the
    International Symposium on Lepton Moments (Cape Cod, June 2003).}
}

\author{Andrzej Czarnecki
\address{Department of Physics, University of Alberta\\
Edmonton, AB\ \  T6G 2J1, Canada}
\\[5mm]
William  J.~Marciano
\address{Physics Department, Brookhaven National Laboratory,
Upton, NY 11973, USA}
\and
Arkady Vainshtein
\address{William I. Fine Theoretical Physics Institute, University of
  Minnesota,\\ 
 116 Church St.\ SE, Minneapolis, MN 55455, USA}}
\maketitle

\begin{abstract}
Hadronic electroweak corrections to the muon anomalous magnetic moment
($g-2$) are reviewed.  Emphasis is on clarification of discrepancies
among various published studies.  A theorem on non-renormalization of
the transversal part of a correlator of two vector currents and an axial
current is reviewed and its consequences in the form of superconvergent 
sum rules are discussed. 
\end{abstract}

\section{Introduction}
\begin{figure}[ht]
\hspace*{38mm}
\begin{minipage}{16.cm}
\psfig{figure=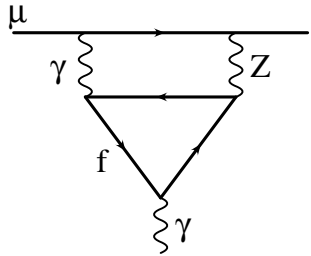,width=40mm}
\end{minipage}
\caption{\sf \small Effective $Z\gamma\gamma^*$ coupling induced by a
fermion triangle, contributing to $a^{\rm EW}_\mu$.}
\label{fig:triangle}
\end{figure}

The subject of this paper is a class of two-loop electroweak contributions to
the muon $g-2$ containing a fermion triangle along with a virtual
photon and $Z$ boson, as
shown in Fig.\,\ref{fig:triangle}.  
In the standard model all charged fermions contribute to the triangle
loop.  Individual contributions of fermions lighter than the $Z$ boson
are enhanced by large logarithms $\ln(M_Z/m)$, where $m$ denotes the
mass of the fermion in the loop or of the muon, whichever 
is heavier.  Those large logarithms were first found in \cite{KKSS}
where the diagrams in Fig.\,\ref{fig:triangle} where evaluated only
for leptons 
in the triangle loop.  However, it was pointed out in \cite{km90} and
explicitly shown in \cite{CKM95} that such logarithms cancel in sums
over all fermions of a given generation (as long as $m\ll M_{Z}$
for all fermions in the generation).  The source of this cancellation
can be traced back to the cancellation of anomalies,
given by the same fermionic triangles, within a given generation.
It is required for renormalizability. 

 Some subtlety here is that the anomaly cancellation refers to the
longitudinal part of the axial $Z$ boson current,  while both
transversal and longitudinal parts of the fermionic triangle contribute
to $g-2$.  At the level of free quarks the cancellation is, of course,
explicit, so the only question is whether it can be spoiled by strong
interactions.  Since $\ln M_Z$ arises from the loop momenta much
larger than the hadronic scale $\Lambda_{\rm QCD}$  perturbative
QCD should provide the answer. For the longitudinal part the
Adler-Bardeen theorem\cite{AB} guarantees the absence of gluonic
corrections governed by $\alpha_{s}(M_{Z})$. We will describe below why
this is also true for the transversal part confirming the cancellation
of $\ln M_Z$ within a generation.

For the light quark loops strong interactions become essential
at the range of loop momenta on the order of hadronic scale. Not only
is 
$\alpha_{s}\sim 1$ but also nonperturbative phenomena defining hadron
masses are crucial in this range. Namely, these dynamical 
phenomena --- not the
current masses of light quarks --- define effective infrared cutoff in
the logarithmic integrals. While an exact calculation in this range is
not possible, in the paper\cite{CKM95} (CKM) a crude approximation was
used: ``constituent masses'' of 300 MeV assigned to quarks $u,d$ and
500 MeV to the quark $s$ played the role of the infrared cutoff.  The
contribution of the first two generations of fermions was found to be
\ba 
\Delta a_\mu^{\rm CKM} (e,u,d; \mu,c,s) &=& {G_F\over \sqrt{2}}
{m_\mu^2 \over 8\pi^2} {\alpha\over \pi} \left[ -6 \ln { (m_u
m_c)^{4/3} \over (m_d m_s)^{1/3} m_\mu^2 } -{49\over 3} + {8\pi^2
\over 9} \right] \nonumber \\[1mm] & \simeq & -8.3 \cdot 10^{-11}
\qquad \mbox{(with $m_c=1.3$ GeV)}.
\label{ckm}
\ea 
  Some uncertainty  was assigned to this result, but the
cancellation of $\ln M_Z$, present in that model, was believed to be
rigorously valid.   

Around the same time the paper \cite{Peris:1995bb} (PPR) appeared, in
which the 
fermion triangle contributions of Fig.\,\ref{fig:triangle} were also
studied.  The result of the PPR paper for the first two generations was
\ba
\Delta a_\mu^{\rm PPR} (e,u,d; \mu,c,s) &=&
{G_F\over \sqrt{2}}
{m_\mu^2 \over 8\pi^2} {\alpha\over \pi}
\left[
-{14\over 3}\ln {M_Z^2 \over m_\mu^2} + 4\ln {M_Z^2 \over m_c^2}
-{107\over 9} + {8\pi^2 \over 9} 
\right]
\nonumber \\[1mm]
& \simeq & -8.7 \cdot 10^{-11}.
\label{ppr}
\ea 
This calculation was done in the chiral limit, so that the $u,d,s$
masses were replaced by $m_\mu$ in \eq{ppr}, just like the electron mass.  
The  $M_Z$ dependence of the logarithms would cancel if the
first coefficient $(-14/3)$ were replaced by $-12/3=-4$.  We have argued
that $-4$ is indeed the correct coefficient and that the factor
$-14/3$ arises due to an incomplete accounting of $u,d,s$ contributions
in Eq.\,(20) in \cite{Peris:1995bb}.   

Since the numerical values of \eq{ckm} and \eq{ppr} are very similar,
and since analytical results in \cite{Peris:1995bb} were given for
individual flavors and not for their sum, the presence of a residual
$\ln M_Z$ in the final result of \cite{Peris:1995bb} was
inconspicuous.  The difference was first pointed out by Mingxing Luo in a
private communication to W.~Marciano.

More recently, the authors of \cite{Peris:1995bb} together with
M.~Knecht have revisited that issue in a detailed study
\cite{Knecht:2002hr}.  They maintained the finding of
\cite{Peris:1995bb} and argued that the cancellation of $\ln M_Z$ in
\cite{CKM95} is a ``spurious'' result of the naive constituent quark
model.  If the conclusion of \cite{Knecht:2002hr} were correct, it
would call into question the validity of QCD studies in a variety of
contexts, since it suggests that low energy strong dynamics may
influence very short distance phenomena, a violation of the basic
tenet of asymptotic freedom.

On the other hand, with respect to large distances, the analyses in
\cite{Peris:1995bb,Knecht:2002hr} correctly emphasize differences in
the hadronic dynamics of longitudinal and transversal parts of quark
triangles. These differences are not properly accounted for by
approximation with constituent masses of light quarks. Moreover, 
constituent masses break the chiral structure of the 
fermion loops.

The above challenge prompted us to reanalyze the triangle
contributions to $g-2$ in \cite{Czarnecki:2002nt}.  We believe to have
located the source of the error in \cite{Knecht:2002hr}; our arguments
are summarized in this text.  We also address some questions raised
after \cite{Czarnecki:2002nt} was published.  Let us stress that the
discrepancy in question is not relevant for an interpretation of the
present $g-2$ experiment, whose design accuracy is $\pm 40\cdot
10^{-11}$.  This is an academic dispute {\em par excellence}.
However, its clarification is important, especially if our community
is to determine the much more difficult hadronic three-loop
light-by-light effects that limit the accuracy of the standard
model prediction for $g-2$.

\section{Hadronic triangle diagrams}
\begin{figure}[ht]
\hspace*{38mm}
\begin{minipage}{16.cm}
\psfig{figure=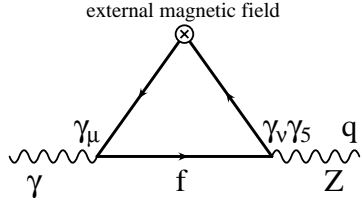,width=50mm}
\end{minipage}
\caption{\sf \small Virtual fermion loop with structure of
  couplings and external momentum.}
\label{fig:anatomy}
\end{figure}

 Some details of the virtual fermion
triangle are shown in Fig.\,\ref{fig:anatomy}.
For the determination of the muon anomalous magnetic moment, we are
interested in the $Z^{*}\to \gamma^{*}$ transition between virtual $Z$
and $\gamma$ in the presence of the external magnetic field to first
order in this field. Moreover, the
magnetic field is constant so $Z^{*}$ and $\gamma^{*}$ carry the same momentum
$q$.  In this approximation, but including all effects of strong
interactions in the quark loop (such as gluon exchanges and
non-perturbative effects), the amplitude of the $Z^{*}-\gamma^{*}$
transition $T_{\mu\nu}(q)$ can be parametrized using two
Lorentz-invariant functions of the external momentum $q$,
\begin{eqnarray}
\label{invfun}
\hspace{-5mm} T_{\mu \nu }(q)&\!\!=\!\!& \frac{e}{4\pi^2}\left[w_T(q^2)\! 
\left(-q^2 \tilde F_{\mu \nu
}\!+\!q_\mu q^\sigma
\tilde F_{\sigma  \nu } \!-\! q_\nu q^\sigma \tilde F_{\sigma  \mu }\right)
\!+w_L(q^2)\, q_\nu q^\sigma \tilde F_{\sigma  \mu }\right]\!.
\end{eqnarray}
Here, $\tilde F_{\mu \nu}=(1/2)\epsilon_{\mu\nu\gamma\delta} F^{\gamma\delta}$ 
denotes the dual of the external
electromagnetic field tensor.  The structure \eq{invfun} is obtained
in the following way.  We construct $T_{\mu\nu}$ which is a
pseudo-tensor under Lorentz transformations using the vector $q_{\mu}$ 
and the pseudo-tensor $\tilde F_{\mu\nu}$.  From these objects we can construct
three  structures: $\tilde F_{\mu \nu}$,
$q_\mu q^\sigma \tilde F_{\sigma \nu}$, and
$q_\nu q^\sigma \tilde F_{\sigma \mu}$.  In addition,  $T_{\mu\nu}$ must
be consistent with electromagnetic current conservation,  $q^\mu
T_{\mu\nu} = 0$.  This leaves us with two independent structures and
we group them in such a way that one is transversal and the other
longitudinal with respect to $q_\nu$.  

The existence of the longitudinal part stems from the axial ABJ
anomaly\cite{ABJ} which in fact exactly determines $w_L(q^2)$ in the
chiral limit of massless fermions $f$, \\[-5mm]
\ba
q^\nu T_{\mu\nu} = {e\over 4\pi^2} w_L(q^2) q^2 q^\sigma \tilde
F_{\sigma \mu}.
\ea
Comparing this with the axial anomaly we find
\ba
\label{wL}
w_L(q^2) = -{2\over q^2} \cdot 2I^{3}_{f}N_{f}Q_{f}^{2}
\equiv {2\over Q^2}\cdot 2I^{3}_{f}N_{f}Q_{f}^{2},
\ea
where $Q_{f}$ and $I^{3}_{f}$ are the electric charge and the third
component of weak isospin projection while $N_{f}$ accounts for the
number of colors in the  case of quarks and equals one for leptons.  
There are no corrections to this
result in the limit $m_f =0$, neither perturbative --- due to Adler-Bardeen
theorem\cite{AB}, nor nonperturbative --- due to 't~Hooft
consistency condition\cite{tHooft}.

There is no such strong result for $w_T$, but there is a very useful
theorem proven by A. Vainshtein \cite{Vainshtein:2002nv}.  It states
that for massless fermions, the following relation holds to all orders
of perturbation theory:
\ba
\label{TL}
2w_T(q^2)=w_L(q^2) \qquad \mbox{(for $m_{f}=0$)}.
\ea
There are nonperturbative corrections to this result which are given
by powers of $\Lambda_{\rm QCD}^{2}/Q^{2}$ but no perturbative ones.
One way to prove Vainshtein's theorem is to consider the imaginary
part of $T_{\mu\nu}$.  The crucial point is that $\Imag T_{\mu\nu}$ is
symmetric under $\mu \leftrightarrow \nu$, $q \leftrightarrow -q$.
Indeed, since $\Imag T_{\mu\nu}$ is given by convergent diagrams, we
can freely use the anti-commutation of $\gamma_5$ to move it from the
axial vertex $\gamma_{\nu}\gamma_{5}$ to the vector one,
$\gamma_{\mu}$.  In the limit $m_f=0$, this involves commuting
$\gamma_5$ with an even number of $\gamma$ matrices, no matter how
many gluon emissions occur on the way.

Thus, we find
\ba
&&\Imag 
\left[w_T(q^2) 
\left(-q^2 \tilde F_{\mu \nu}
+q_\mu q^\sigma \tilde F_{\sigma  \nu } 
- q_\nu q^\sigma \tilde F_{\sigma  \mu }\right)
+w_L(q^2)\, q_\nu q^\sigma \tilde F_{\sigma  \mu }\right]
\nonumber \\[1mm]
&&\qquad 
\propto 
q_\mu q^\sigma \tilde F_{\sigma  \nu } 
+
q_\nu q^\sigma \tilde F_{\sigma  \mu }. 
\ea
This is possible only if 
\ba
\label{delta}
2\,\Imag w_T=\Imag w_L={\rm const} \,\delta(q^{2}).
\ea
Only with such a form for $\Imag w_{T,L}(q^{2})$, $\Imag T_{\mu\nu}$
has no terms antisymmetric in $\mu \leftrightarrow \nu$.  The
coefficient of the delta-function in Eq.\,(\ref{delta}) is fixed to be
$2\pi\cdot 2I^{3}_{f}N_{f}Q_{f}^{2}$ by the exact form  of
$w_{L}$ in (\ref{wL}). Both $w_T(q^{2})$ and $w_L(q^{2})$ are
analytical functions decreasing at large $q^{2}$. It means that they
satisfy unsubtracted dispersion relations
\ba
\label{disp}
w_{T,L}(q^{2})=\frac{1}{\pi}\int_{0}^{\infty}\!\!{\rm d} s\,
\frac{\Imag w_{T,L}(s)}{s-q^{2}} \,,
\ea
which when combined with  Eq.\,(\ref{delta}) for imaginary parts
implies relation (\ref{TL}) for the real parts as well. 

Note that the real part of $T_{\mu\nu}$ contains an antisymmetric term
$\tilde F_{\mu\nu}$.  In terms of diagrammatic calculations it arises
from counterterms fixed by the conservation of the vector
current. When the Pauli-Vilars regularization is used these
counterterms are given by heavy regulator loops.

To summarize, we now know that in the chiral limit $m_{f}=0$, the
longitudinal function $w_{L}$ is exactly given by Eq.\,(\ref{wL}) and
the transversal function $w_{T}$ is known up to nonperturbative
corrections,
\ba
w_T = {2I^{3}_{f}N_{f}Q_{f}^{2}\over Q^2}
 + \mbox{non-perturbative corrections}
\label{wT}
\ea
At this point we can apply our findings to an analysis of logarithmic
contributions $\ln M_Z$ to $g-2$.  We will use the fact that
Eqs.\,\eq{wL} and \eq{wT} rigorously determine the large $Q^2$
asymptotics of $w_{L,T}$; the non-perturbative corrections to $w_T$,
to be discussed below, are suppressed by extra powers of $\Lambda_{\rm
QCD}^2 / Q^2$ and do not influence $\ln M_Z$  terms. Corrections due to
deviations from the chiral limit are also suppressed by powers of
$m_{f}/Q$.

How do $w_L$ and $w_T$ enter the $g-2$ contribution of the diagrams in
Fig.~\ref{fig:triangle}?  It was shown in \cite{Czarnecki:2002nt} that
for a determination of $\ln M_Z$ terms we can use the following simple
representation:
\begin{equation}
\label{amu}
\Delta a^{\rm EW}_\mu
\simeq \frac{\alpha}{\pi }\,\frac{G_\mu\,m_\mu ^2}{8\pi^2\sqrt{2}}
\int_{m_\mu^2}^\infty {\rm d}Q^2 \left(w_L+\frac{m_Z^2}{m_Z^2+Q^2}\,w_T
\right).
\end{equation}
Two points can be made regarding this integral:
\begin{itemize}
\item
$\int^\infty {\rm d}Q^2 w_L$ diverges.  The theory is inconsistent
unless the anomaly cancellation condition is satisfied,
\begin{equation}
\label{cancel}
\sum_f I^3_f \,N_f\,Q_f^2=0
\label{anomcan}
\end{equation}
\item
\vspace{-2.5mm}
$\int^\infty {\rm d}Q^2M_{Z}^{2}\,w_T/(m_Z^2+Q^2)\simeq \ln M_Z$.
With $w_{T}=w_{L}/2$ at large $Q$ the anomaly cancellation condition
\eq{cancel} leads to the cancellation of $\ln M_{Z}$ within a
generation where $m_{f}\ll M_{Z}$ for all fermions.
\end{itemize}

In the analyses \cite{Peris:1995bb,Knecht:2002hr} the authors missed
the leading $1/ Q^{2}$ part in $w_T(u,d,s)$ but not in
$w_L(u,d,s)$ and $w_{T,L}(e,\mu,c)$.  Naturally, they arrived then
at the expression \eq{ppr} where $\ln M_{Z}$ terms do not cancel.
Technically the mistake in \cite{Peris:1995bb,Knecht:2002hr} arose
from an incorrect construction of the Operator Product Expansion (OPE)
in the part related to $w_{T}$. Referring to \cite{Czarnecki:2002nt}
for details note that the authors missed the leading operator which
reflects interaction with the soft electromagnetic field (external
magnetic field) at short distances. As we demonstrated above this
leads to an inconsistency.

On the other hand, the authors of \cite{Peris:1995bb,Knecht:2002hr}
were correct in their analysis of nonperturbative corrections, they
were the first to point out that these corrections start with
$\Lambda_{\rm QCD}^{4}/Q^{6}$ in $w_{T}$. So, below we will try to
explain following\cite{Czarnecki:2002nt} what can be done when we
account for both the leading perturbative term $1/Q^{2}$ in $w_{T}$
and subleading nonperturbative terms $\Lambda_{\rm QCD}^{4}/Q^{6}$ and
higher.

\section{Large $Q^2$ expansion, sum rules and models for $\Imag w_{T,L}$}
\begin{figure}[ht]
\hspace*{0mm}
\begin{minipage}{16.cm}
\begin{tabular}{c@{\hspace{16mm}}c}
\psfig{figure=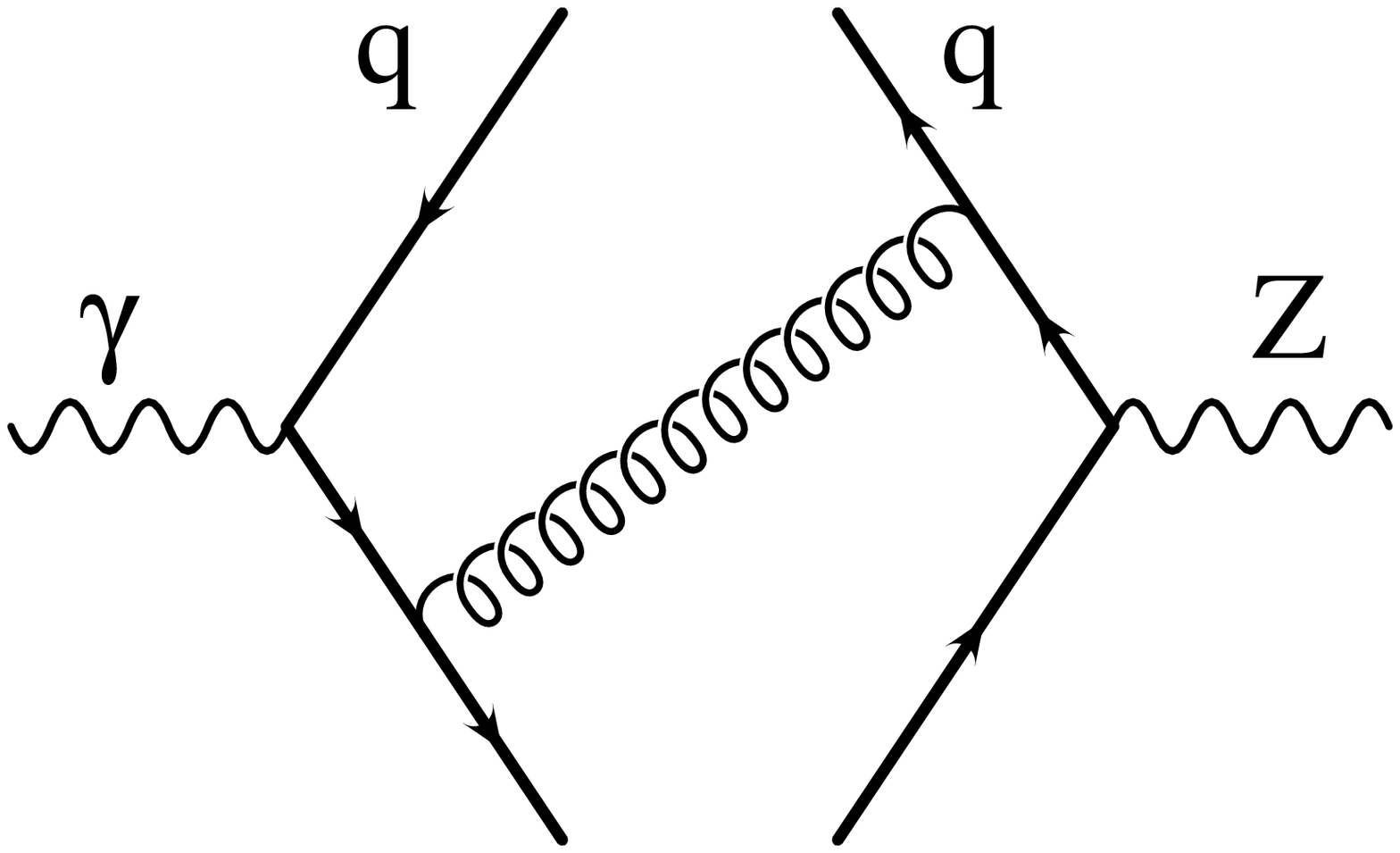,width=48mm}
&
\psfig{figure=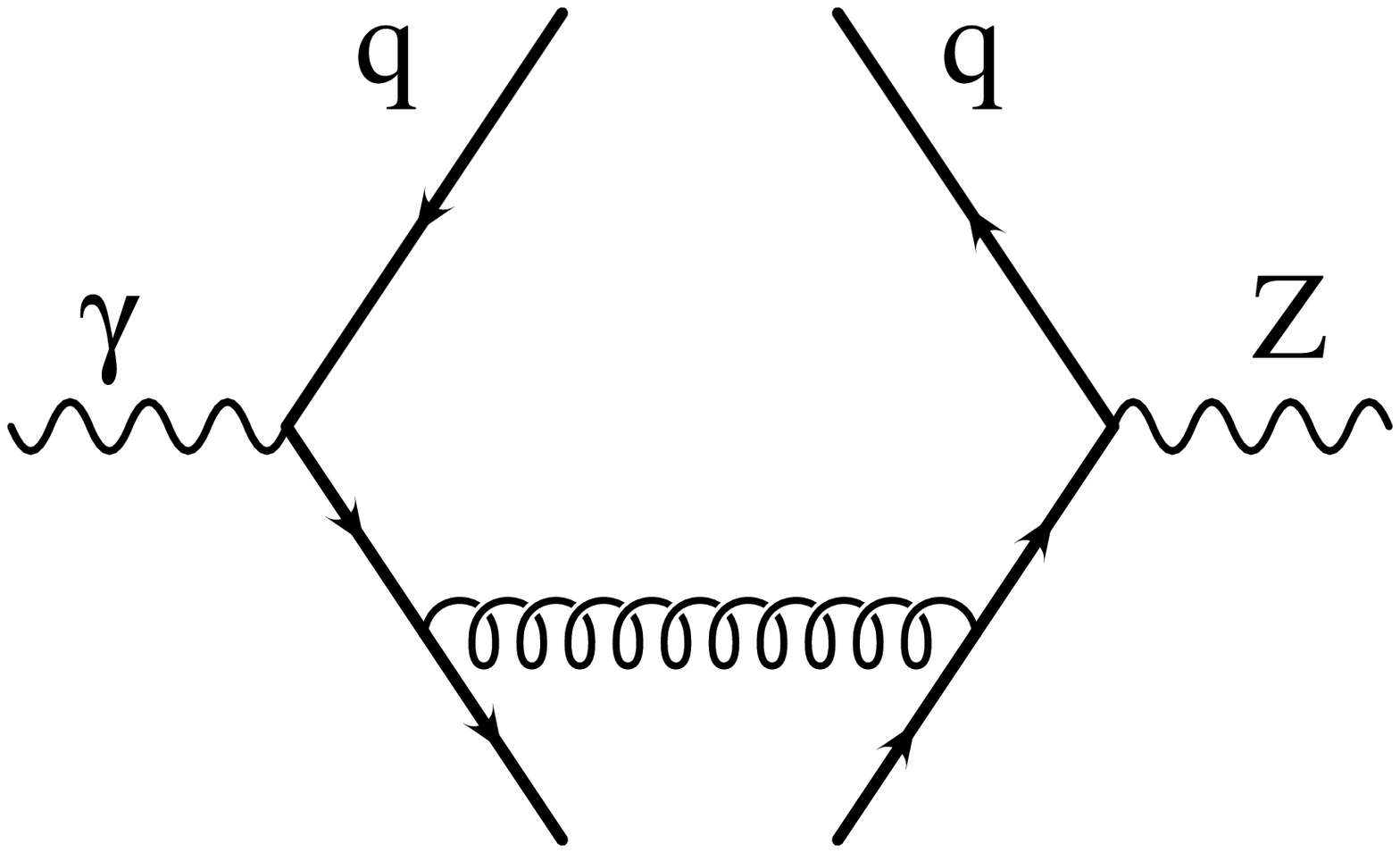,width=48mm}
\end{tabular}
\end{minipage}
\caption{\sf \small Diagrams for four-fermion operators responsible for
  leading non-perturbative corrections to $w_T$.}
\label{fig:fourF}
\end{figure}

For simplicity we limit ourselves in this section to the first
generation, i.e.~$u$ and $d$ quarks.  As we mentioned above, the
first nonperturbative corrections in the chiral limit $m_{u,d}=0$ are
of order $\Lambda_{\rm QCD}^{4}/Q^{6}$ in $w_{T}$, as has been shown
in detail in \cite{Knecht:2002hr,Czarnecki:2002nt}. A particular
example illustrating how the $d=6$ operators appear in the OPE is
given in Fig.~\ref{fig:fourF}.
The diagrams allow for a perturbative calculation of the OPE
coefficient while averaging of the four-fermion operator in the
external magnetic field involves nonperturbative physics.

Of course, the expansion of $w_{T}$ in powers of $1/Q^{2}$ continues
further and we are not able to find the sum.  Still the analysis gives
some exact relations: the first term is $1/Q^{2}$ with the
coefficient 1 (it is normalized to $\sum_{u,d}
2I_{3}N_{f}Q_{f}^{2}=1$), while the  coefficient of $1/Q^{4}$ is zero.
These relations can be rewritten in the form of sum rules for
$\Imag w_{T}$. Indeed, using the dispersion representation \eq{disp} at
large $Q^{2}$ we find
\be
\label{srwt}
\int_{0}^{\infty}\!\!{\rm d} s\, \Imag w_{T}(s)=\pi\,, \qquad
\int_{0}^{\infty}\!\!{\rm d} s \,s \,\Imag w_{T}(s)=0\,. 
\ee
The existence of such sum rules has been pointed out in
\cite{edrCCtalk}, where their exactness was considered puzzling.  It
does not look puzzling to us, as there are many examples of such exact sum
rules like Weinberg sum rules or sum rules for the DIS
structure functions.  Moreover, even if $w_T$ would not have the
leading $1/Q^2$ term, as assumed in \cite{Knecht:2002hr}, these sum
rules would be no less exact (except the right hand side would be zero
in the first one).

It is instructive to compare these relations with similar sum rules
for the longitudinal function $w_{L}$ where all powers of $1/Q^{2}$
are known, 
\be
\int_{0}^{\infty}\!\!{\rm d} s\, \Imag w_{L}(s)=2\pi\,, \qquad 
\int_{0}^{\infty}\!\!{\rm d} s\,s^{k} \,\Imag w_{L}(s)=0\quad (k=1, 
\ldots, \infty)\,.
\ee
This set of relations implies a unique solution $\Imag w_{L}(s)=2\pi
\delta(s)$ showing that the massless pion is the only contributing
intermediate state.  

For the transversal function $w_{T}$ the intermediate hadronic states
have to be $1^{+}$ mesons with isospin 1 and 0 or $1^{-}$ mesons with
isospin 1. The lightest ones $\rho$, $\omega$ and $a_1$ are massive
even in the chiral limit. 
Representing $\Imag w_{T}$ as 
\be
\Imag w_{T}=\pi \sum_{i} g_{i} \delta(s-m_{i}^{2})
\ee
we get from Eq.\,\eq{srwt} 
\be
\label{sres} 
\sum_{i} g_{i} =1\,,\qquad \sum_{i} g_{i} \,m_{i}^{2}=0\,.
\ee
The analysis of \cite{Knecht:2002hr} assumed $\sum_{i} g_{i} =0$.

Until now we did not make any specific assumptions.  
Let us now assume a saturation by the lightest states and consider
$\rho$ and $\omega$ to be degenerate. Then the solution of relations
\eq{sres} gives 
\be
 w_{T}=\frac{1}{m_{a_{1}}^{2}-m_{\rho}^{2}}
\left[\frac{m_{a_{1}}^{2}}{Q^{2}
+m_{\rho}^{2}}-\frac{m_{\rho}^{2}}{Q^{2}
+m_{a_{1}}^{2}}\right]
\ee
This model for $w_{T}$ by construction satisfies exact
relations \eq{srwt} and was used in \cite{Czarnecki:2002nt}. Of
course, one can modify the model adding extra resonances but
numerically $a_{\mu}$ is not much sensitive to these modifications.

One more comment on the chiral symmetry breaking by quark
masses. Accounting for nonvanishing $m_{u,d}$ leads to negligible
effects for $w_T$. The main (and important) effect for $w_{L}$ is a
shift of the pole from 0 to $m_{\pi}^{2}$, i.e. to
$w_{L}=2/(Q^{2}+m_{\pi}^{2})$.

These refinements of $w_{T,L}$ in the long distance hadronic triangle
effects for the $u,d$ quark loops, together with similar ones for the
$s$ quark, were made in our study \cite{Czarnecki:2002nt}.
Numerically it changes the result in Eq.~\eq{ckm} to $\Delta
a_\mu^{\rm CMV} (e,u,d; \mu,c,s) = -6.7(1.0) \cdot 10^{-11}$.

\section{Summary}
Although the combined hadronic and leptonic triangle loop effects in
electroweak corrections are not that 
significant for the muon $g-2$, it is reassuring that a theory based
on the Operator Product Expansion and its matching with hadronic
phenomenology allows for quite accurate calculations. We resolved a
conceptual controversy existing in literature concerning cancellation
of short distances between fermionic triangles.  It seems to be
possible now to develop an analogous approach to improve theoretical
accuracy for the light-by-light part of hadronic effects.

\section*{Acknowledgments}
This research was supported in part by the Natural Sciences and
Engineer\-ing Research Council of Canada and by the DOE grants
DE-AC03-76SF00515 and DE-FG02-94ER408.

\end{document}